
\input harvmac
\Title{\vbox{\baselineskip12pt\hbox{LA-UR-92-3477}\hbox{hep-th/9301060}
}}{Generating Vector Boson Masses}
\centerline{Amitabha Lahiri\footnote{$^{\dag}$}{(lahiri@pion.lanl.gov)}}
\bigskip\centerline{Theoretical  Division  T-8}
\centerline{Los Alamos National Laboratory}
\centerline{Los Alamos, NM 87545, USA}
\vskip0.3in
\centerline{\bf Abstract}

\noindent  If the Higgs particle is never found, one will need an
alternative theory for vector boson masses. I propose such a theory
involving an antisymmetric tensor potential coupled to a gauge field.

\Date{\vbox{\hbox{10/92}}}
\def\sb{$SU(2)\times U(1)$}
\def\Bmn{B_{\mu\nu}}
\def\Frl{F_{\rho\lambda}}
\def\Fmn{F_{\mu\nu}}
\lref\sgr{S. G. Rajeev, {\sl Duality and Gauge Invariance},
MIT preprint CTP-1335.}
\lref\abl{T. J. Allen, M. J. Bowick and A. Lahiri, {\sl Mod. Phys. Lett.}
{\bf A6} (1991) 559;  {\sl Phys. Lett.} {\bf B237} (1989) 47.}
\lref\mw{J. Minahan and R. Warner, {\sl
Stuckelberg Revisited},\ University of Florida preprint UFIFT-HEP-89-15.}
\lref\kr{S. Deser, {\sl Phys. Rev.} {\bf 187} (1969) 1931\semi M. Kalb and P.
Ramond,
{\sl Phys. Rev.} {\bf D9} (1974) 2273\semi R. Rohm and E. Witten, {\sl Ann.
Phys.} {\bf 170} (1986) 454.}
\lref\leblanc{M. LeBlanc {\it et al},  {\sl Mod. Phys. Lett.} {\bf A6}
(1991) 3359.}
\lref\ws{S. Weinberg, {\sl Phys. Rev.} {\bf D19} (1979)	1277\semi
L. Susskind, {\sl Phys. Rev.} {\bf D20} (1979) 2619.}
\lref\al{A. Lahiri, {\sl An Alternative Scenario for Non-Abelian Quantum Hair},
LA-UR-92-471, hep-th@xxx/9202045, {\sl Phys. Lett.} {\bf B} (in press).}
\lref\ttbar{Y. Nambu, Enrico Fermi Institute preprints 88-39(1988), 88-62(1988)
and 89-08(1989)\semi W. A. Bardeen, C. T. Hill and M. Lindner, {\sl Phys.
Rev.} {\bf D41} (1990) 1647.}
\lref\velt{M. J. G. Veltman, {\sl Acta Phys. Polo.} {\bf B8} (1977) 475;
{\sl Phys. Lett.} {\bf B70} (1977) 253.}


The Standard Model of electroweak interactions is, as its name implies, a
well-established theory both from theoretical and experimental viewpoints.
However, there are still a few unanswered questions about it. Perhaps the
biggest of them is about the Higgs sector of the theory. The basis of the
misgivings is that no elementary scalar has ever been found, and there is
no positive evidence at currently available energies of a Higgs particle.
On the other hand various analyses set the upper bound of the Higgs mass
only a little out of our current reach.
This suggests that perhaps it is time we prepared ourselves for
the situation that no Higgs particle is ever found within the predicted
range of energies.

What is the purpose of the Higgs sector? It is intimately related to
spontaneous symmetry breaking. It breaks the global electroweak \sb\
symmetry, thus generating masses for vector bosons. It also contributes to
the mass generation mechanism for fermions via chiral symmetry breaking.
There have been various attempts
to treat these effects without the use of an elementary scalar particle.
For example, in technicolor theories \ws, one postulates a new strong
interaction which binds techniquark-antiquark pairs into a chiral
condensate that breaks the electroweak symmetry. In another scenario
\ttbar, one uses the top quark rather than techniquarks to form
condensates. Both these routes try to replace the Higgs sector by something
else. In this letter I make a similar attempt and propose an alternative
mechanism for generating vector boson masses. This mechanism is somewhat
similar in its aspects to the theory of a heavy Higgs particle \velt, since
the abelian theory is dual to an abelian Higgs model with infinitely
massive Higgs. However, any connection between the theory proposed here and
that of heavy Higgs bosons (or that of nonlinear $\sigma$-models) is yet to
be discovered.  This letter is intended to be a digest of results; details
will be published elsewhere.

The proposed alternative involves the non-abelian generalization of the
theory of a two-form potential \kr\ interacting with a gauge field. The
abelian theory has been known to render the photon massive \refs{\abl,\mw}
and has the added virtue of not having a residual degree of freedom (\`a la
the Higgs particle). Attempts at producing massive non-abelian gauge bosons
this way have come up short of expectations because of the complexity and
difficulty involved in non-abelianizing the vector gauge symmetry
associated with the two-form (see, for example, \sgr\ and
\leblanc). In what follows I offer
solutions to some of the well-known problems with this system.

{\sl The Unadorned Theory:}
The starting point of this theory is a naive non-abelianization of the
 mass generation mechanism \abl\ for electrodynamics via a $B\wedge F$
coupling between a two-form $\Bmn$ and the field-strength $\Fmn$ of the gauge
field. One adds a kinetic term for $\Bmn$ and the usual $F^2$ to get the
desired result. (For simplicity I do not include fermionic matter in this
model.) To move on to an $SU(N)$ gauge group the derivative operator $\del_\mu$
is replaced by the covariant derivative $D_\mu$ and $\Fmn = [D_\mu, D_\nu]$,
resulting in the action
\eqn\lgf{S = \int d^4x\Tr\bigg(-{1\over 6}H_{\mu\nu\lambda}H^{\mu\nu\lambda} -
{1\over 4}\Fmn F^{\mu\nu} + {m\over
2}\epsilon^{\mu\nu\rho\lambda}\Bmn\Frl\bigg), }
where $H_{\mu\nu\lambda} = D_\lambda\Bmn + cyclic$, with $\Bmn$ in the adjoint
representation of $SU(N)$. The classical equations of motion derived from this
action are invariant under an additional global field redefinition $\Bmn
\to \Bmn + \alpha\Fmn$, where $\alpha$ is a constant. For the moment I
exclude other terms from the action by requiring this invariance. (It is
possible that this invariance will be broken in the quantized theory. That
however does not affect the applicability of this theory to
mass generation, as I will discuss below.)

Firstly I shall discuss the attractive features of this theory. The
quadratic part of this action is exactly what is needed to give the gauge
boson a tree level mass. This mass is equal to $m$ and occurs as a pole in
the gauge boson propagator upon summing all tree level two point diagrams
with insertions of $\Bmn^a$ propagators ({\it i.e.}, by diagonalizing the
fields at zero gauge coupling). The classical counting of modes is as
follows --- if I denote the conjugate momenta to $A_\mu$ and $\Bmn$
respectively by $\Pi^\mu$ and $\Pi^{\mu\nu}$, the constraints are
\eqn\constraint{\eqalign{\Pi^0 \approx 0,\qquad \Pi^{0i} &\approx 0,\hfill\cr
D_j\Pi^j + [B_{ij}, \Pi^{ij}] &\approx 0, \hfill\cr
D_j\Pi^{ij} + {m\over 2}\epsilon^{ijk}F_{jk} &\approx 0.\hfill\cr }}
This forms a closed, reducible system of constraints. The closure is easily
verified using Poisson brackets $\{A_i, \Pi^j\} = \delta^j_i, \{B_{ij},
\Pi^{kl}\} = \half(\delta^k_i\delta^l_j - \delta^k_j\delta^l_i)$. (The
factor of $\half$ arises from a conflict in conventions, while
$D_{[\mu}D_{\nu]} = D_\mu D_\nu - D_\nu D_\mu, B_{[\mu\nu]} = \Bmn \equiv
\half(\Bmn - B_{\nu\mu})$.)  If I define
$b_i := \half\epsilon_{ijk}B_{jk}$ and the conjugate momenta $\Pi^{(b)}_i$, the
last set of constraints are equivalent to the abelian
$\vec\nabla\times\vec\Pi^{(b)} + {m
\over 2}\epsilon^{ijk}F_{jk}\approx 0$. This shows that the constraints are
not independent and one needs to fix a gauge, say $\nabla^jB_{ij} = 0 =
\vec\nabla\times\vec b$. This shows that there is a surviving longitudinal mode
coming from the
$B$ field. This is of course the mode that couples with the gauge field to
produce a massive boson.

Obviously, the masses of the gauge bosons are all equal in this model. If
this is to be applicable to electroweak phenomena, $m$ has to be replaced
by a mass matrix. One takes the viewpoint that the symmetry breaking is
induced by some mechanism not yet included in this model (possibly by
couplings to fermions), that selects out a $U(1)$ group from \sb. Since the
end result should be one massless photon and three massive gauge bosons,
this is equivalent to diagonalizing a $4\times 4$ mass matrix $m_{ab}$
(over the \sb\ Lie algebra), which has exactly one vanishing eigenvalue.
Diagonalization leads to the difference between $W^{\pm}$ and $Z$ masses
and gives the Weinberg angle. Finally, this model is power-counting
renormalizable and also seems to be renormalizable at the one-loop level,
as I point out in the next section.

{\sl Symmetries:}
The principal obstacle to writing a dynamical theory of a non-abelian two-form
is the lack of symmetries. While the abelian theory has a vector gauge symmetry
under $\Bmn \to \Bmn + \del_{[\mu}\Lambda_{\nu]}$, this symmetry does not seem
to survive non-abelianization. In the absence of this symmetry, the constraints
\constraint\ (specifically the last one of the set) are very difficult to
implement on the fields, and questions arise concerning the renormalizability
and unitarity of the model. This situation is similar to quantizing a
spontaneously broken gauge theory in the unitary gauge.
The main point of this letter is to propose that this symmetry is actually
hidden
in the theory given by \lgf. I introduce an auxiliary field $C_\mu$, also in
the adjoint representation of the gauge group, and rewrite the action as
\eqn\auxac{S = \int d^4x\Tr\bigg(-{1\over 6}(H_{\mu\nu\lambda}
- [F_{[\mu\nu},C_{\lambda]}])(H^{\mu\nu\lambda} - [F^{[\mu\nu}, C^{\lambda]}])
- {1\over 4}\Fmn F^{\mu\nu} + {m\over
2}\epsilon^{\mu\nu\rho\lambda}\Bmn\Frl\bigg).}
%
This action now has two gauge symmetries
\eqn\gauge{\eqalign{A_\mu &\to UA_\mu U\inv - \del_\mu UU\inv,\hfill\cr
\Bmn &\to U\Bmn U\inv, \hfill\cr
C_\mu &\to UC_\mu U\inv;\hfill\cr}}
and
\eqn\KR{\eqalign{A_\mu &\to A_\mu,\hfill\cr
\Bmn &\to \Bmn + D_{[\mu}\Lambda_{\nu]}, \hfill\cr
C_\mu &\to C_\mu + \Lambda_\mu, \hfill\cr}}
where $\Lambda_\mu$ is any vector field transforming in the adjoint
representation. The action \lgf\ is then seen as the gauge-fixed ($C_\mu =
0$) version of this action. Since the action \auxac\ contains no quadratic
term for $C_\mu$, one needs to introduce one in order to compute Feynman
diagrams of the theory. This is most simply done
with a fake mass term $\Tr({1\over 2\eta}C_\mu C^\mu)$.
One then takes $\eta \to 0$ after calculating with a given cutoff. This
decouples $C_\mu$ from the theory and diagrams containing internal $C_\mu$
propagators vanish. This is good from the point of view of
renormalizability of the theory. The gauge field sector is renormalizable,
of course, and one needs to consider only the sector involving the $\Bmn$
field. The couplings appearing in the action \lgf\ are all dimension four,
propagators for both $A_\mu$ and $\Bmn$ fall off as $1/k^2$ at large
momenta, and it follows (and can be explicitly checked) that possible
counterterms appearing at the one loop level are at most dimension four.
The renormalization of this theory is quite involved and I shall discuss it
in more detail elsewhere. The dimensionality of the counterterms leads one to
hope that the theory is renormalizable even when the global symmetry is broken.
It also turns out that the relevant Ward
identity for the two-point function of $\Bmn$ implies that only the
longitudinal mode of the $\Bmn$ field is affected by higher order
corrections. As a result one should expect $\Bmn$ to remain
longitudinal at higher orders in perturbation theory. Therefore, even if
higher order diagrams break the
global symmetry, they will not produce a Higgs particle, nor invalidate this
theory as a mechanism for generating vector boson masses. More specifically, no
counterterm of the form $B^2$ or $B\wedge B$ arises as higher loop corrections.
Finally, one hopes
that the symmetries \gauge\ and \KR\ in the action
\auxac\ are enough to decouple non-propagating modes from the theory. While
that seems obvious at the classical level, it needs to be verified after
quantization.

{\sl Comments and Summary:}
There are various open problems associated with this theory. One of the more
important ones is the question of couplings of $\Bmn$ to fermions. Since the
theory with $C_\mu = 0$ does not have the vector gauge symmetry
\KR, fermions may couple to $\Bmn$ via couplings either of the form
$\psibar\sigma^{\mu\nu}\Bmn\psi$ or
$\psibar\gamma^5\sigma^{\mu\nu}\Bmn\psi$. (One replaces $\Bmn$ by $(\Bmn -
D_{[\mu}C_{\nu]})$ in order to see the gauge symmetry \KR\ explicitly.)
These couplings then obey the gauge symmetries but break chiral symmetry.
One expects contributions to magnetic (or electric) dipole moments of
fermions from such couplings.

In the abelian version of this theory there is no residual degree of
freedom, {\sl i.e.}, there is no Higgs particle. Analysis of the non-abelian
theory points to the same conclusion. However, the possibility of a
condensate corresponding to the Higgs particle cannot be ruled out.

It is tempting to apply this theory to QCD, as the aforementioned global
transformation allows one to remove a $\Theta F\wedge F$ term from the
Lagrangian, even for arbitrarily small values of the bare mass parameter
$m$.  However, it is not clear if this actually solves the strong CP
problem, or just sweeps it under the rug.

I have not ascertained the full implications of this theory on the
phenomenology of electroweak or strong interactions. It seems that the most
immediately obvious impact will be on processes that involve the production or
exchange of Higgs bosons in the standard model. Tree level processes should
otherwise be indistinguishable form those in the standard model. One should
also expect differences arising from the possible fermion couplings mentioned
above.

In conclusion, I have proposed in this letter a mechanism for generating masses
for non-abelian vector bosons that apparently does not have an excitation
corresponding to the Higgs particle. \sb\ symmetry breaking can also be
incorporated into this model without invoking a Higgs particle. This model also
seems to be renormalizable, and should be of interest to field theorists and
phenomenologists alike --- to theorists because of the rich structures in both
the classical and quantum theories, and to phenomenologists because processes
corresponding to this model differ from the standard model predictions and can
be tested in high-energy experiments. In the event that the Higgs particle is
never found, the model proposed here should be a viable alternative for the
Higgs sector of the standard model.

\medskip
\vbox{\centerline{\bf Acknowledgements}

It is a pleasure to acknowledge long, insightful discussions with A.
Kovner.  I also thank T. Allen, M. Bowick, N. Dorey, M. Mattis and E.
Mottola for discussions at various stages of this work.}

\listrefs\bye